\documentstyle[12pt,aaspp4]{article}

\begin{document}

\title{Deep CCD Surface Photometry of the Edge-On Spiral NGC 4244}
\author{Anne M. Fry}
\affil{Department of Astronomy, Case Western Reserve University, 
Cleveland OH 44106-7215 
\\ electronic mail: anne@smaug.astr.cwru.edu}
\author{Heather L. Morrison\footnote{Cottrell Scholar of Research
Corporation and NSF CAREER fellow}}
\affil{Department of Astronomy and Department of Physics, 
Case Western Reserve University, Cleveland OH 44106-7215 
\\ electronic mail: heather@vegemite.astr.cwru.edu}
\author{Paul Harding}
\affil{Steward Observatory, University of Arizona, Tucson, Arizona 85726
\\ electronic mail: harding@as.arizona.edu}
\and
\author{Todd A. Boroson}
\affil{US Gemini Program, National Optical Astronomy 
Observatories\footnote{The National Optical Astronomy Observatories 
are operated by the Association of Universities for Research in Astronomy, 
Inc (AURA) under cooperative agreement with the National Science
Foundation}, Tucson, Arizona 85726\\ 
electronic mail: tyb@noao.edu}

\begin{abstract}
We have obtained deep surface photometry of the edge-on spiral galaxy
NGC~4244. Our data reliably reach 27.5 $R$ magnitude arcsec$^{-2}$, a
significant improvement on our earlier deep CCD surface photometry of
other galaxies. NGC~4244 is a nearby Scd galaxy whose total luminosity is 
approximately one magnitude fainter than the peak of the Sc luminosity 
function. We find that it has a simple structure:  a single exponential 
disk, with a scale height $h_Z$ = 246 $\pm$ 2 pc, a scale length $h_R$ = 1.84 
$\pm$ 0.02 kpc and a disk cutoff at a radius $R_{max}$ = 10.0 kpc (5.4 scale lengths).
We confirm a strong cutoff in the stellar disk at $R_{max}$, which happens over 
only 1~kpc. We do not see any statistically significant evidence for disk flaring 
with radius. Unlike the more luminous Sc galaxies NGC~5907 and M~33, NGC~4244 does
not show any evidence for
a second component, such as a thick disk or halo, at $\mu_R < 27.5$ magnitude arcsec$^{-2}$.

\end{abstract}

\section{INTRODUCTION}

In order to better characterize the distribution of luminosity in the
stellar components of disk galaxies, we are conducting a detailed
study of edge-on disk galaxies with a range of bulge-to-disk ratio
using deep CCD surface photometry.  Our galaxies range in Hubble type
from S0 to Scd, and in bulge-to-disk-ratio from 20 (NGC 3115) to 0.03 
(NGC 4244, this work \& \cite{bbj98}). We will re-examine the structure of their 
bright disks and bulges, and also study, for the first time, the structure of faint 
outer components such as thick disks and stellar halos. 

These faint outer components are important because they
may trace the history of the galaxy, either because they contain the
oldest stars (\cite{e62}, \cite{e93}) or because they were
formed via accretions of smaller galaxies (\cite{sz78}, \cite{d85},
\cite{qg86}).  In addition, if they are made of old stars, they should
provide connections to high-z galaxy observations, both direct
(\cite{stei}) and indirect via absorption line studies (\cite{wolfe90}).

The simplest galaxies structurally are those of late type -- Sc and
later.  To first order, they comprise simply a disk with exponential
surface brightness falloff (\cite{freeman70}, van der Kruit \& Searle 
1981\nocite{vdks}).  We have already studied a large Sc galaxy, 
NGC~5907 (Morrison et al. 1994\nocite{mbh}). Our analysis confirmed previous 
findings on its thin disk (van der Kruit \& Searle 1981\nocite{vdks}), but we 
found an unusual second component, whose luminosity falls off very slowly with 
radius (see \cite{sack94}, \cite{jm98}).  Since the galaxy luminosity function 
(LF) for Sc galaxies is very broad, and peaked at total blue luminosity 
($M_{B,tot}$) of --19.5 (\cite{bst88}), making NGC 5907 ($M_{B,tot} = -20.55$, 
\cite{tully}) an unusually luminous Sc galaxy, we have also studied NGC~4244, a
smaller late-type spiral whose total magnitude of --18.4 $B$ mag
(\cite{o96a}) is almost a magnitude below the broad peak of the field
Sc LF.  NGC~4244 has Hubble type Scd, is located 3.6 Mpc away
(Tully-Fisher, \cite{a86}), and has a flat rotation curve with $V_{rot}
= 100$~km/s (\cite{o96a}). Beyond the optical disk, NGC~4244 has a declining rotation
curve (\cite{o96a}).

In this paper we present deep surface photometry of NGC~4244. Our data
reach reliably to $\mu_R$ = 27.5 magnitude arcsec$^{-2}$, significantly fainter
than our previous published data on NGC 5907. Questions that we will
address using this extremely deep data are:

\begin{itemize}
\item do small late-type disk galaxies such as NGC 4244 have similar 
properties to large ones such as NGC 5907?

\item what are the properties of NGC 4244's luminous thin
disk?  Does the stellar disk have a distinct edge?  What is its
vertical scale height?  Does this scale height remain constant with
radius? 

\item does NGC 4244 show fainter components like a thick disk or halo, or like
NGC~5907's unusual second component?

\end{itemize}

NGC~4244 is less massive than a typical Sc:
the total mass of NGC~4244 is 9.77 x 10$^9 M_{\sun}$ (\cite{tully}), whereas
the average total mass of an Sc is $\approx 10^{11} M_{\sun}$ (\cite{rharaa}).
However the H~I mass fraction $M_{H I}/M_{tot}$ (\cite{o96a})
is fairly typical for an Sc galaxy: 0.14. The mean $M_{H I}/M_{tot}$ for a typical Sc 
(from \cite{rharaa}) is 0.1. The mass fraction of H$_2$ ($M_{H_2}/M_{tot}$) is
low in NGC~4244, relative to other galaxies of the same mass.  
In his CO study, \cite{sage93} found $<M_{H2}/M_T> = 0.0013$, which placed 
NGC~4244 in his mass class III, for which  $<M_{H2}>$ = 8.4 x 10$^7 M_{\sun}$ or
$<M_{H2}/M_T> \approx 0.008$. 
The current star formation rate in NGC 4244 is also 
low, as can be seen from its low X-ray, radio continuum, and IRAS emission (cf. 
\cite{o96a}). 

The Hubble type of a galaxy is known to correlate roughly with its
environment in the sense that late-type galaxies are found in more
isolated areas (\cite{dress80}).  In addition, we might expect to see
a correlation between the existence of thick disks and galaxy
environment if the accretion theory of thick disk formation is
correct.  NGC~4244 lies in the Coma-Centaurus spur, and is thought to
be a member of the CVn~I (\cite{dv75}) or B4 (\cite{kkt79}) group. Its
small distance makes group membership very difficult to decide, since group
peculiar velocities are of the same size as the Hubble flow. We have plotted in Figure \ref{enviro}
supergalactic coordinates of galaxies in the B4 and nearby B5 groups of Kraan-Kortweg \& Tammann -- it is not clear whether
these two groups are separate entities. We can only conclude from these
data that NGC~4244 lives in a relatively isolated environment, with only three galaxies of
similar or greater brightness within $\sim$~2 Mpc (NGC~4449, NGC~4736, NGC~4826),
the brightest of which (NGC~4736) has M$_{B,tot}$ = $-$19.37 mag, 1.7 mag brighter than
NGC~4244.

\cite{vdks} used photographic surface photometry of NGC~4244
to study its structure, finding the disk well modeled by a radial
exponential function and a vertical sech$^2$ function, with a sharp
cutoff at 5.3 radial scale lengths.  They found no evidence for a thick
disk, in contrast to their studies of disk galaxies with
larger B/D ratios where thick disks were seen.  However, thick disks
are very faint, compared to both the night sky and the light of the younger, thin disks,
and, thus, detection and accurate determinations of the properties of thick disks from photographic work are
difficult.  The improved precision of CCDs is needed, and our faint
CCD photometry of NGC 4244 will put much stronger limits on the
existence of a thick disk in this galaxy.

The flaring of a galaxy's stellar disk gives constraints on the
balance between the gravitational potential the disk experiences and
heating mechanisms the stars experience.  van der Kruit \& Searle 
(1981a,b,1982a,b\nocite{vdks}\nocite{vdks81b}\nocite{vdks82b}\nocite{vdks82a}) 
found that stellar disks in the galaxies they studied had constant scale heights with
radius, although this has recently been challenged (de~Grijs \& Peltier 1997\nocite{dgp}).  
NGC~4244 is a good galaxy in which to study disk flaring, as
it has no large companions and no apparent thick disk
(cf. van der Kruit \& Searle 1981a\nocite{vdks}). We will discuss the 
strong limits  our data place on any variation of disk scale 
height with radius.

\section{OBSERVATIONS \& REDUCTIONS}

The NGC~4244 data were obtained over a five night run in March 1997
and a six night run in April 1997 using the 24\arcsec /36\arcsec ~ Burrell Schmidt
telescope\footnote{Observations made with the Burrell Schmidt of the
Warner and Swasey Observatory, Case Western Reserve University.} at
Kitt Peak National Observatory\footnote{Kitt Peak National Observatory
is a division of the National Optical Astronomy Observatories (NOAO),
which are operated by the Association of Universities for Research in
Astronomy, Inc. (AURA) under cooperative agreement with the National
Science Foundation}. The images were taken using a 2048x2048
frontside-illuminated Tektronix CCD (S2KA) mounted at Newtonian
focus. With this setup, the field of view is 69 square arcmin, with
each 21 $\mu$m pixel imaging 2.$^{\arcsec}$03 of sky. The gain was set
to 2.5 e$^-$/ADU and the readout noise was 3 e$^-$ (1.2~ADU). All
exposures were made through a Harris $R$ filter.

Observations were performed in the manner described by
\cite{m97}. Half of the time was used observing the galaxy and the other half 
was used to obtain dark sky flats. The dark sky flat images were
taken at approximately the same hour angle and declination as the
galaxy images. Over the course of the two observing runs, 66 galaxy images
and 70 dark sky flats were obtained. For all of the galaxy and sky
images, the exposure time was 600~s. Conditions were photometric
throughout the observing runs.

Overscan removal and bias subtraction were done in the standard manner. Then
``master'' sky flats were constructed from the dark sky images from
each run. The individual sky flats were dithered by $\sim5^{\circ}$, so
that wings from bright stars in the field would not appear in the
final, combined image.  To construct the master sky flat, we used an
iterative procedure.  We scaled the individual sky images by their
mode. Prescaling was important because, even during photometric
conditions, the sky brightness varied by $\sim$0.3 mag. This was
likely due to the long tail to high wavelength of the Harris (Cousins)
$R$ filter, which includes some of the strong airglow lines, combined with the change
in sky brightness with zenith distance. Then, we
combined the individual, scaled, sky images, using IRAF's ccdclip
algorithm, which uses CCD properties to estimate expected variations
in data values and removed pixels which differed from the median by
more than 2$\sigma$. Each of the individual sky frames was divided by
the flatfield frame to reduce the width of the distribution of sky
values and so make rejection of outliers due to faint stars and
stellar wings more accurate. A plane was fit to each of the individual
flat-fielded sky frames and subtracted from the sky frame. The modes
were recalculated, and the entire procedure repeated. This procedure
was repeated until no improvement was seen in the master sky flat (ten
times in this case).

The 66 galaxy images were dithered by $\sim400^{\arcsec}$,
much less than for the sky flats. The individual galaxy images were 
flatfielded using the master sky flat and registered. We combined the 
individual galaxy images, using a median. In Figure \ref{final} we show 
our final galaxy image, after subtracting a best fit plane sky and
correcting to airmass 1.

One Landolt standard star field (\cite{l92}) was observed each night,
giving us 26 well-exposed standard stars with a range in color and
airmass. A photometric zero point of $R$ = 22.23 $\pm$ 0.03 magnitude arcsec$^{-2}$
 corresponding to 1 ADU/sec/pixel was determined. For a 600 s
exposure, this yields $R$ = 29.18 $\pm$ 0.03 magnitude arcsec$^{-2}$
corresponding to 1 ADU/exposure/pixel. As our exposures were only taken
in one filter, we cannot add a color term to our calibration but
estimate its magnitude as less than 0.10 mag over the color range of our standards
(see Figure \ref{stds}, $B-V$ = $-$0.06 to 1.66 mag, $V-R$ = 0.05 to 0.8).
In fact, over the expected color range of the relatively old populations we 
will be studying, the color term is well below the uncertainty in the zero point.

The average sky brightness, 1998.0 ADU / pixel (see Section 3.2), corresponds to
$R_{sky}$ = 20.9 magnitude arcsec$^{-2}$.

\section{ANALYSIS \& RESULTS}

For simplification in subsequent analyses, the galaxy has been rotated
counter-clockwise by 42$^{\circ}$. This agrees well with Olling's (1996a)\nocite{o96a}
determination of $-$48$^{\circ}$ for the position angle of NGC~4244.

\subsection{Masking \label{ming}}

In order to reach the faintest possible surface brightness level in
the galaxy, we masked out the foreground stars and background galaxies in
our field. To characterize the stars we used DAOphot (\cite{s87}) to
determine the positions and magnitudes of all of the stars within the
frame. Using 48 bright, but unsaturated, stars, we determined the
point spread function (PSF) for stars in the frame. We followed the
method described in Stetson (1987)\nocite{s87}, iterating until a
clean subtraction within a 15x15 pixel box centered on the PSF star
was obtained.  
Our aim was not to determine accurate photometry of all
of the stars within the frame, but to mask the stars out to a constant
level in surface brightness. This means that we paid particular
attention to the outer wings of our PSF. We used this PSF
to construct a radial profile and masked all pixels for which a
foreground star contributed at least 1 ADU. The stellar radial profile
was well determined out to the PSF radius, 15 pixels (30$\arcsec$).
The PSF is axisymmetric and varies along the chip. The variation is 
most significant for the central regions of the stars and does not
strongly affect the outer wings of the stars, which are formed by
processes such as scattering in the atmosphere and telescope.
We describe the PSF beyond 15 pixels in Section 3.4.

Background galaxies and undetected stars were masked by hand using the
IRAF task imedit.  Special care was taken with stars near the galaxy,
where the galaxy's luminosity gradient  interferes with
DAOphot's sky fitting algorithm.

NGC~4244 has a small dust lane and several visible \ion{H}{2}
regions. To mask these, a dust mask 20 pixels wide, running along
the major axis, was created. In Figure \ref{maskd} we show the galaxy
with our final mask superimposed.

Using this mask, the data were binned up into regions whose size varied from
a resolution element near the galaxy's center to 50~x~50 pixels in the regions
far from its center. This procedure used a robust averaging scheme to 
compensate for any unmasked stars, galaxies, etc. and is described in detail
in \cite{mbh}.

\subsection{Sky Subtraction\label{skysub}}

Accurate sky subtraction is crucial for the faint isophotes. To
determine the sky level, we used the masked image and made a histogram of 
sky values in bins well away from the galaxy, using bins with more than 100 pixels
after masking. There are 310 such bins. Using a
histogram of the sky values within these bins, we determined the sky
value to be 1998.0 $\pm$ 0.8~ADU. In Figure \ref{histsky}, we show this histogram.
The width of the histogram provides a measure of the contribution of unmasked stars
to the sky level.

\subsection{Limits to Our Precision}
Our photometry reaches reliably to 27.5 magnitude arcsec$^{-2}$.  The error
model is described in Appendix A. 
To illustrate, we work through the errors in a 13 x 3 pixel
bin, centered 12 pixels (425 pc) from the major axis and 150 pixels
(5.3 kpc) from the minor axis. In this bin, the mean number of counts
is $\overline C = 2498$ ADU per pixel, with the mean number of counts from the
galaxy alone $\overline g = 500$ ADU per pixel.  The errors are summarized in
Table \ref{galerr}. The dominant source of error in the thin disk
region of the galaxy is the error arising from surface brightness
fluctuations within NGC~4244, because NGC 4244 is so close to the
Milky Way (its distance is 3.6 Mpc; note that it is marginally
resolved even from the ground.) Surface brightness fluctuations
dominate the formal errors wherever $\overline g / n_{pixels} > 0.40$
(i.e. up to 1.6 kpc above the plane). In the faintest region of the
galaxy (above 1.6 kpc from the plane), the main sources of error are
large-scale flat-fielding errors and sky variations.  

\begin{dummytable} \label{galerr} \end{dummytable}

\subsection{Scattered Light?}

To see how much scattered light from the central part of the galaxy contributes 
at high-z, we constructed a stellar profile out to 150 pixels (5.3~kpc).
In the center, we used the stellar profile constructed using DAOphot (Section \ref{ming}).
For the outer region, we used two saturated stars. Using stars whose centers were saturated 
means that the signal-to-noise in the outer wings is higher than in unsaturated stars. We masked 
the stars in the wings of the saturated stars, as well as diffraction spikes around the saturated 
stars. The unmasked 
pixels were averaged in radial annuli. The profiles from the two saturated stars were averaged 
and joined to the DAOphot PSF profile. The combined profile was normalized to the surface 
brightness of the central bright region of NGC~4244. The final profile is shown in
Figure \ref{psf}. At a distance of 2~kpc (56~pixels) from galaxy center, the scattered
light from the bulge is $\approx$~28.5 magnitudes arcsec$^{-2}$, below our quoted photometric limit
of $\mu_R = 27.5$~ magnitudes arcsec$^{-2}$. Thus, it is not a significant effect.

\subsection{2-D Fits}

We fit a single exponential disk to the data:
\begin{eqnarray}
\rho_{disk} = L_0 e^{-R/h_R} e^{-|z|/h_z}~~~(R < R_{max});\\
\hspace{0.1in}= 0~~~(R \geq R_{max}.)
\label{rho}
\end{eqnarray}
(here $L_0$ is the central density of the disk in $L_{\sun}$~pc$^{-3}$).
The two-dimensional modeling is described fully in \cite{mbh}. We start with a full 
three-dimensional model, and then for each iteration of the fit, we
numerically integrate along the line of sight and produce a two-dimensional model, which we compare with the
observed surface brightness.

We start out with eight parameters: the scale length and scale height
$h_R$ and $h_Z$, inclination $i$, the coordinates of the galaxy center
$x_C$ and $y_C$, the radial extent of the galaxy $R_{max}$, the
central surface brightness $I_C$, and the sky value. Because of high
correlations among the parameters, eight free parameters are extremely 
difficult to fit accurately, but, fortunately, some of the parameters can be 
fixed using other information.  The sky value is determined by averaging the values in
any bin with more than 100 pixels, away from
the galaxy, as described in Section 3.2. The $x_C$ and $y_C$ position of the center of the galaxy
and $R_{max}$ are determined from a contour plot of the galaxy and
by making one-dimensional profiles across the minor axis and parallel
to the major axis. The inclination $i= 84.5^\circ$ was adopted from
Olling's (1996a\nocite{o96a}) kinematic measurement.  Only three
parameters then need to be determined from the 2-D fitting of a single
exponential disk: $h_R$, $h_Z$, and central surface brightness. 

How well do 2-D fits work?  Because we are using a Levenberg-Marquardt algorithm (\cite{nr}), 
which is an iterative nonlinear leastsquares algorithm, we need to
provide starting values.  We investigated the sensitivity of the
algorithm to the initial estimates used by mapping out the value of
$\chi^2$ over a large grid of starting values and noting where the
algorithm converged. We find that if all three parameters, $h_R$,
$h_Z$, and $I_C$, are estimated, then the initial estimates
need to be within 20\% of the correct values for the model fit to
converge. If one of the three parameters can be estimated from other
information, such as $h_Z$ from the 1-D fits, then initial estimates
of the other two can be as much as 50\% wrong and the model will still
converge successfully. A starting grid of $\chi^2$ can be used as a
``Levenberg-Marquardt-by-eye'' to locate the ``trench'', that is, the
starting parameter values for which the model fit will converge. Once
the trench has been found, all three parameters are allowed to vary.

As illustration, in Figure \ref{trench}, we show a contour plot of
$\chi^2$ for fixed $h_Z$ = 6.5 pixels (233 pc). The contours are
logarithmic; steep slopes are even steeper than they appear. For all
starting values of $h_R$ and $I_C$ shown on the plot, except upper
left ($h_R \approx 1000, I_C \approx 110$), lower left ($h_R \approx 1000, 
I_C \approx 30$), and lower right ($h_R \approx 3000, I_C \approx 30$) 
corners, the starting model will converge. However, for $h_Z$ values 
which are more than $\approx 20$\% from the true $h_Z$, the initial
values of $h_R$ and $I_C$ need to be within $\approx$50\% of the
true values.

Another concern is the reliability of the estimates once the
algorithm has converged.  \cite{mbh} showed that estimates based on the
covariance matrix of the fit can underestimate the true values
severely. Thus, to determine error estimates on the derived disk parameters $h_R$,
$h_Z$, and $I_C$, we used bootstrapping with resampling (\cite{e82}, Chapter 15 of 
\cite{nr}, \cite{et93}). 
We ran our 2-D model fitting program 1000 times. Each time,
the data to be fit were chosen by random sampling with replacement from the actual data.
The sets of resulting disk parameters were roughly Gaussian distributions,
centered on the result of the run using all of the data. The errors estimated
from the widths of bootstrapping analysis are more reasonable
estimates of our true errors, as the bootstrapping errors include not
only the formal errors from known contributions but also a more realistic estimate of
goodness of fit and systematics (unmasked foreground stars, dust, and
small, faint, background galaxies). In Figure \ref{hist}, we show a
histogram of $h_Z$ values estimated by our bootstrapping code.

In Table \ref{twod}, we show the results of our 2-D fitting with the bootstrapping
errors, as well as the correlation coefficients between all the variables.
To determine all of the correlation coefficients, all eight variables
were started at their correct values and allowed to vary. The important
correlation coefficients are those between the first three variables,
the scale length $h_R$, the scale height $h_Z$, and the central surface 
brightness $\mu_0$\footnote{Note that the central surface brightness quoted here is 
for the observed inclination of 84.$^{\circ}$5. For comparison to other galaxies, the 
face-on and edge-on central surface brightnesses for this model are also quoted.} But 
the strong correlations between some of the other 
variables, such as between $h_Z$ and inclination, or $R_{max}$ and $h_R$,
demonstrate why it is important to use other methods to estimate some
of the parameters. \begin{dummytable} \label{twod} \end{dummytable}

We estimated $R_{max}$ directly from the image to reduce the effects of these correlations.
Unfortunately, the galaxy is not axisymmetric:
$R_{max}$ on the southwest side of NGC~4244 (left side of Figure 4) is 8\% larger than on
the northeast side: 10.5~kpc vs 9.7~kpc. We chose an intermediate value of 10.0~kpc. 
Using the northeast side (right side of Figure 4) only in our estimation, with 
$R_{max}$~=~9.7~kpc, results in an estimate of $h_R$ which 
is 3\% larger. The estimates of $h_Z$ and $\mu_0$ change an insignificant amount. The bright stars 
and \ion{H}{2} regions at the left edge of NGC~4244 make it difficult to model the left side only.
  
\subsection{Vertical Profiles \label{oned}}
In Figure \ref{profiles}, we show vertical profiles parallel to the
minor axis. Our best fit model of a single exponential disk with scale height of
246 pc is shown with the data. A single disk fit is a very good match
to the data down to 27.5 magnitude arcsec$^{-2}$ and there is little evidence for
disk flaring, i.e. changing scale height as a function of radius. The only
profile that appears to flare is the one at $\pm$~8.5 kpc. However, the 
warp in NGC~4244 starts at $\approx~8.0$ kpc and we attribute the apparent
flaring to the warp. In the galaxy center, we may have a small contribution to
the light from the bulge, which may lead to a slight underestimate of $h_Z$ there, due
to the steeper bulge profile.

To test the constancy of $h_Z$ with radius, we split the galaxy into
inner and outer regions, $R < 4$~kpc and $4 < R < 8$~kpc (the region outside
$R = $8~kpc was not included to minimize the effect of the small warp), and analyzed
the two regions separately. For the inner region, we find $h_Z$ = 246 $\pm$ 2 pc
and, for the outer region, we find $h_Z$ = 259 $\pm$ 8 pc. The difference 
between the inner and outer regions is not significant at the 2$\sigma$ level. This 
lack of evidence for flaring will be discussed further in Section 4.3.

In Figure \ref{xp} and \ref{xp2}, we show profiles parallel to the major axis, at $\pm$ 760~pc
and at $\pm$ 955~pc from the major axis. The horizontal profile at z~=~$\pm$ 760~pc is
the closest distance to the major where a complete profile can be drawn. This is because of the
large number of pixels masked due to dust, \ion{H}{2} regions, and foreground stars at lower
z. In Figure \ref{xp}, the region further out than $\sim$ 7~kpc from
the minor axis shows more light than predicted by a single exponential disk fit, as if either the outer
regions have a larger scale length $h_R$ or the inclination changes slightly with radius, becoming
closer to edge-on for large $R$. 
Olling's\nocite{o96a} (1996a) \ion{H}{1} data also suggest that
the inclination increases slightly with radius.

The profile at z~=~$\pm$ 955~pc shows the influence of the warp more clearly.
Olling (1996a)\nocite{o96a} found evidence for a warped \ion{H}{1} disk. 
He found that the warp starts close to the edge of the optical
disk ($R$~=~8.5~kpc), as is typical of galaxy warps (\cite{briggs}). The \ion{H}{1} warp
extends below the major axis on the left side of the disk and 
above on the right side in our Figure 3. \cite{o98} finds that its line of nodes is 
close to the line of sight. In Figure \ref{xp2}, the warp is apparent for the entire outer 
half of the disk, starting at
5~kpc from the minor axis, where data from the four quadrants of our image are quite 
different. In other words, in our faint optical data we see more details of the warped outer disk
superimposed on the line of sight than is seen in the \ion{H}{1} data, despite its high quality. 
(We do not show data past 7~kpc on the left side because of the presence of
a large region of current star formation and three bright stars.) 

We also see evidence for a small warp in the outer parts of the stellar
disk in our vertical profiles. The line of centers in the vertical 
profiles changes by $\approx$2.5 pixels (89~pc) between the
center and the left edge of the disk. We corrected for this effect before making Figure 9.
This is equivalent to a change
of 0.$^{\circ}$5. The optical warp we find is in the same direction
as the \ion{H}{1} warp. The optical warp in NGC~4244 is much smaller than in
NGC~5907 (Morrison et al. 1994\nocite{mbh}) or NGC 4565, which permits 
accurate study of its outer stellar disk.

\section{DISCUSSION}

\subsection{Comparison to Previous Surface Photometry}

\cite{vdks} used photographic photometry to obtain thin disk
parameters for NGC~4244 and, recently, Olling (1996a)\nocite{o96a} reanalyzed their
photometry to rederive a disk scale height and disk scale length.
We confirm the existence of a disk cutoff, first noted by  \cite{vdks},
with our deeper data. \cite{ky96} used $J$ and $K^{\prime}$ photometry to estimate 
thin disk parameters. We list our disk parameters with those of the other groups
in Table \ref{diskcmp}, including the orientation (``as-observed'', face-on, and edge-on). 
\cite{vdks} adopted a distance of 5 Mpc, rather than 
3.6~Mpc used by the other studies, and, in Table \ref{diskcmp}, we have scaled 
their scale height and scale length to the distance adopted by the other studies.
To compare their $B_J$ data to our $R$ photometry, we assume $B-B_J = 0.2$ and 
$B-R$~=~1.5, colors typical for an old stellar population (\cite{b89}). 
\cite{vdks} fit their data to a sech$^2$ model, rather than an exponential
light distribution and we include an estimate of the central surface brightness from an
exponential model: $\mu$ (exponential) $= \mu$ (sech$^2$) - 0.75 mag. 

IF we compare our $R$ photometry to van der Kruit \& Searle's\nocite{vdks} $B_J$
photometry, we find a blue color ($B-R$ = 0.2~mag, rather than 1.5~mag) for NGC~4244.
However, it is very difficult to determine accurate zero points from photographic photometry;
$B$ CCD data are needed to determine the galaxy's color.

Our disk scale height is 15\% larger than found by \cite{vdks} and by Olling 
(1996a)\nocite{o96a}, but our scale length is the same. Our disk parameters do not 
agree well with those found by \cite{ky96}, however, their data are limited to 
the inner $180\arcsec$ (3.2~kpc), making the scale length more difficult to
determine. In summary, our results agree well with previous optical studies.
\begin{dummytable} \label{diskcmp} \end{dummytable}

In Table \ref{galcmp}, we compare our disk parameters for NGC~4244 to
those of NGC~5907 (Morrison et al. 1994\nocite{mbh}) and the Milky Way (cf. 
Morrison et al. 1997\nocite{m97}). NGC~4244 has a smaller disk scale height 
than the other two galaxies and a fainter central surface brightness. The ratio 
of disk cutoff to radial scale length is similar to that of the Milky Way.
\begin{dummytable} \label{galcmp} \end{dummytable}

\subsection{Disk radial cutoff}

\cite{vdks} showed that stellar disks do not stretch out indefinitely,
showing a sharp cutoff at $\sim$5 radial scale lengths, well inside the \ion{H}{1} disk. Our deep
surface photometry of this almost unwarped stellar disk gives a
strong test for disk cutoffs. Figure \ref{cutoff} shows profiles just inside $R_{max}$
($<$~10 kpc) and outside it (11.3 kpc).  All the points outside $R_{max}$ have 
$\mu_R$ fainter than 27 magnitude arcsec$^{-2}$. The
cutoff in stellar light happens relatively rapidly, over a distance of
1 kpc, and no faint outer component is visible down to
$\mu_R$= 27.5 magnitude arcsec$^{-2}$. The existence of this stellar cutoff is consistent with
the detection by \cite{ferg98} of faint \ion{H}{2} regions ``at
distance twice the optical radius'' of the nearby face-on galaxies they studied. The
``optical radius'' (corresponding to a face-on surface brightness
$\mu_B$ = 25 magnitude arcsec$^{-2}$) is an outmoded concept based on what could be detected
on photographic (POSSI) sky survey plates; it is now possible to see
much fainter with CCD detectors. (For edge-on galaxies the
situation is, of course, even easier.) Assuming $h_R$/$h_Z$ = 10 
to convert face-on to edge-on surface
brightness, $\mu_B$ = 25 corresponds to $\mu_R \sim$ 21. Figure
\ref{profiles} shows that much of the stellar luminosity of NGC
4244 comes from outside the ``optical radius'' defined in this way.

\subsection{Constant Scale Height}

What determines the thickness of a stellar disk? Since its brightness varies 
exponentially with radius and the 
contribution of dark matter should not be strong in the inner galaxy, 
the weaker gravitational pull from the outer disk means that we would 
expect a component whose vertical energy does not vary with radius to flare
towards the edges.  This can be seen clearly for NGC 4244's \ion{H}{1} in
Fig. 13 of Olling (1996a)\nocite{o96a}, the scale height of
\ion{H}{1} increases from $\approx400$ pc at a radius of 5 kpc to 1.5
kpc at 13 kpc.  However, the factors which
determine the scale height of a {\it stellar} disk are more complex:
although stars are born out of ISM with low velocity dispersion which
is constant with radius (\cite{bronf}, \cite{wouter}, \cite{sco}, \cite{combes}), 
various heating mechanisms change the velocity dispersion and, thus, the 
scale height. The velocity dispersion of stars in the solar neighborhood 
increases with age (\cite{jw83}). This is thought to be due to heating by 
spiral arms and giant molecular clouds, with GMCs being particularly important 
for vertical heating (\cite{jb90}).  van der Kruit \& Searle found that the
scale height of the stellar disks in the galaxies they studied (1981a,b, 
1982a,b\nocite{vdks}\nocite{vdks81b}\nocite{vdks82a}\nocite{vdks82b})
remained constant with radius, despite the variation in CO surface density and,
so presumably, numbers of GMCs with radius.

Recently, \cite{dgp} claim evidence for disk flaring in their study of
48 edge-on galaxies, and for a trend of increasing flaring with Hubble type
(in the sense that earlier types show more flaring). For S0 galaxies 
(de~Vaucouleurs type $\approx~-$2), they predict a change in scale height
with radius of  $\approx 9-10\%$.
For Sc galaxies (de~Vaucouleurs type $\approx$ 5), they predict a change in 
scale height with radius of $\approx 1-4\%$, which is consistent with
our result of a statistically insignificant variation between the 
inner and outer parts of the disk for NGC~4244. It is hard to understand this
variation with Hubble type if the flaring is caused by the factors discussed above.
de~Grijs \& Peltier fit only regions between 1.5 and 4.0 scale lengths, so the 
gravitational effect of the bulge can be ignored in their study.

For earlier type galaxies, however, the determination of disk flaring is 
more difficult: as noted by \cite{dgp}, galaxy interactions, warps, and thick
disks can also cause an apparent change in scale height. An incorrect bulge-disk
decomposition will have the same effect. To distinguish
flaring from the presence of a thick disk requires fainter surface photometry
than a survey project like de~Grijs \& Peltier's\nocite{dgp} can achieve.
Viewed from outside, the Milky Way's thick disk would be visible at $\mu_I$~=~23
magnitude arcsec$^{-2}$ (\cite{m99}, assuming $R-I$~=~0.5), which is  the surface brightness level at 
which de~Grijs \& Peltier's\nocite{dgp} single disk vertical profiles (their Figure 3) begin 
to deviate from a single exponential fit. 

As discussed in Section \ref{oned}, the adopted inclination, which can be an indication
of a warp, can have a large effect on apparent flaring. Kinematic inclination determinations,
such as Olling's for NGC~4244\nocite{o96a}, do not exist for many galaxies, making the 
detection of some warps and correct modeling of the disks more difficult.
de~Grijs \& Peltier do not address the presence of warps in the galaxies they observed. 

In conclusion, the question of how the thickness of stellar disks vary with radius is a 
thorny one to approach observationally, especially in early type galaxies where the
existence of additional components like bulges and thick disks makes accurate separation of
the thin disk light very difficult (see Morrison et al. 1997\nocite{m97} for a discussion
of how the problem is exacerbated if one-dimensional profiles are used instead of 
two-dimensional fits). Our data show a statistically insignificant increase in scale height 
between the inner and outer parts of the disk of NGC~4244, and it is possible that this is 
partially due to the small warp. This highlights the need for more theoretical work 
on disk heating mechanisms. 

\subsection{Is there a second component?}

A single exponential disk (Equation \ref{rho}) is a good fit to the
data down to 27.5 magnitude arcsec$^{-2}$ (see Figure
\ref{profiles}). We do not see evidence of an extra component. 
However, our data may not rule out a faint component resembling the Milky Way's halo, which 
has a very centrally concentrated space distribution, with density $\rho$ varying as $r^{-3}$
or $r^{-3.5}$. NGC~4244 has a nucleus with total luminosity 2\% of its disk's, which we
calculate using the disk central surface brightness and bulge effective intensity given in 
Baggett et al. 1998\nocite{bbj98},
and a bulge effective radius of 4\arcsec (71~pc)\footnote{Baggett et al. give a bulge effective
radius of 400\arcsec (7.1 kpc). We assumed that this was actually 4\arcsec (71 pc).}.
What if this nucleus was the central part of such a halo? We assume that the core radius of this 
putative halo is two pixels (71~pc)\footnote{the core radius of the Milky Way's halo is not
well determined but is probably less than 500~pc based on counts of RR~Lyraes in bulge fields},
and calculate what brightness a power-law halo which fit smoothly onto the luminosity
distribution of the nucleus would be. On the minor axis, 2~kpc above the plane, the
$r^{-3}$ halo would have a brightness $\mu_R$ = 26.5 magnitude arcsec$^{-2}$, which our data 
rule out, but the $r^{-3.5}$ halo would have a brightness $\mu_R$ = 28.4 magnitude arcsec$^{-2}$, below our detection limit.

It is interesting to compare NGC~4244 with its Local Group analog, M~33, which has a 
bulge-to-disk ratio of $\sim$0.02 (\cite{bothun92}). While it is difficult to search for 
halo field stars in M~33 because of its nearly face-on inclination and badly warped disk,
there is evidence for a small number of globular clusters with halo kinematics. \cite{schommer}
show that M~33 has $\sim$20 old clusters in a kinematically hot configuration (see also
\cite{ata}). Whether these clusters are better associated with M~33's disk or tiny bulge,
their kinematics suggest that any field stars associated with them would have a larger 
scale height than the thin disk.

There have been no searches for globular clusters in NGC~4244, but we see no evidence for an 
extended field star population in this galaxy. A globular cluster search in NGC~4244 would 
help us decide whether it truly has no halo.

NGC~5907, a more luminous Sc, has a second component, which does not resemble either the
Milky Way's halo or thick disk (\cite{sack94}, Morrison et al. 1994\nocite{mbh}). 

\section{CONCLUSIONS}

We have obtained deep $R$ band surface photometry of the small, edge-on Scd
galaxy NGC~4244. 
\begin{enumerate}
\item We find that NGC~4244 is well-described by a single
exponential disk with a scale height of 246 $\pm$ 2 pc, a scale
length of 1.84 $\pm$ 0.02 kpc, and a radial cut-off at 5.4 disk scale lengths.
The thin disk parameters, $h_R$, $h_Z$, and $R_{max}$ for 
NGC~4244 are smaller than those of NGC~5907 by about 50\% (cf. Morrison et al.
1994\nocite{mbh}). 
\item NGC~4244 is a good galaxy for investigating stellar disk flaring 
because it does not have a significant thick disk and because it
has no large companions. We find, in agreement with \cite{vdks}, statistically 
insignificant flaring in the stellar disk. The warp in NGC~4244 may exaggerate the marginal 
disk flaring that we see. 
\item We have placed stronger limits on the existence of a cutoff in the stellar disk.
\item Unlike the more luminous Sc galaxies M~33 and NGC~5907, NGC~4244 does not show a significant 
second component. 
\end{enumerate}

\appendix
\section{The Error Model}

It is necessary to have accurate error estimates for both the 1-D and
2-D fitting. Unlike earlier photographic work, deep CCD surface
photometry allows us to quantify measurement errors.  Measurement
errors arise from CCD behavior such as readout noise and flat-fielding, as well
as from intrinsic galaxy variations. Each error contribution will be
addressed below.

\subsection{Readout Noise}
The readout noise per exposure is 1.2 ADU. By combining 66 galaxy images 
with a median, we are able to reduce the effective readnoise to
\begin{equation}
R_{eff} = 1.2~\rm{ADU}~ \frac{1.22}{\sqrt{N_G}} = 0.18~ADU.
\end{equation}
Averaging the 66 frames would have reduced $R_{eff}$ to 0.15 ADU; however, 
combining with a median is preferable for the removal of cosmic rays and other data 
defects. The factor of 1.22 is because of the reduced efficiency of a median (see 
Morrison, Boroson \& Harding 1994\nocite{mbh}).

\subsection{Photon Noise}
For $C$ ADU in a given pixel, the photon noise is $\sqrt{C/g}$, where $g$ 
is the gain. Combining 66 galaxy images using a median reduces the photon 
noise to  
\begin{equation}
\sigma_{Poisson} = \frac{1.22}{\sqrt{66}} \frac{\sqrt{C}}{\sqrt{g}}. 
\end{equation}
In the final combined galaxy image, the contribution to the error from photon 
noise is 0.137~$\sqrt{C}$~ADU.

\subsection{Flatfielding Errors}
In principle, the only limit to the precision of the combined flat field image 
is the photon noise in the individual flat field images. This small-scale 
variation is 
\begin{equation}
\sigma_{sff} = \frac{\sqrt{C_S}}{\sqrt{g}} \frac{1.22}{\sqrt{N_f}} \frac{1.22}{\sqrt{N_G}},
\end{equation}
where $C_S$ is the number of counts in the final, combined master sky flat 
image, $g$ is the gain, $N_F$ is the number of individual sky flats used to 
make the master sky flat, and $N_G$ is the number of individual galaxy images 
used to make the final galaxy image.  The sky counts $C_S$ = 1998.0 ADU, the 
gain is 1.2 ADU, and the numbers of sky and galaxy images are 70 and 66, 
respectively.

In practice, the small-scale flat fielding errors are not the only flat-fielding 
error we have. There are also large-scale variations which arise from the 
variation of the sky brightness across an image and from the wings of bright 
stars which have not been completely removed by combining the individual
sky flats. To estimate the large scale variations $\sigma_{lff}$, we divided 
our sky flats into two subsets of 12 each and constructed two mini-master sky 
flats. We then used IRAF's blkavg to bin the mini-masters into 100x100 pixel 
bins, to remove small-scale variations, and divided one mini-master
into the other. The standard deviation of the divided images was 0.085\% 
$C_S$. The large-scale variation $\sigma_{lff}$ is reduced by $1/\sqrt{2}$ 
because the mini-masters were constructed from half as many sky images as 
the master and by another 1/2 because of the division of the two mini-masters.
The large scale flat-fielding variation $\sigma_{lff}$ = 0.03\% $C_S$. 

\subsection{Surface Brightness Fluctuations}
Another source of error arises from intrinsic variations in the galaxy. The 
light in each pixel comes from individual stars, and, thus, is subject to 
counting statistics. The variance of the brightness fluctuations is 
characterized by \cite{ts88}. Following their Equation 10 and \cite{mbh}, 
we used $t = 600s$, $d = 3.6$ Mpc, $\overline{M} = 0$ (for the $R$ band), 
and $m_1 = 22.23$ magnitude arcsec$^{-2}$. The variance of the brightness is 
3.61$\overline{g}$, where $\overline{g}$ is the number of counts from the
galaxy only.

Seeing has the effect of lessening the apparent brightness fluctuations. We 
modeled the effect of seeing empirically, by creating an IRAF image with 
known $\sigma_{orig}$ and smoothing the image by a Gaussian of similar FWHM 
as our typical seeing. The smoothed image was binned in bins of 3x3 to 50x50 
pixels and $\sigma$ recalculated for each bin size. Without the effect of 
seeing, $\sigma_{orig}$ would be reduced by a factor of $1/\sqrt{n}$, where 
$n$ is the bin size. We calculated the residuals 
$\Delta \sigma = \sigma - \sigma_{orig}/\sqrt{n}$ for each bin size. A function 
was fit to the $\Delta \sigma$ residuals and this function was used to correct 
the surface brightness variation errors for seeing.

\subsection{Sky Variations}
Our error in determining the sky level was 0.8 ADU, or $R$ = 29.3 magnitude 
arcsec$^{-2}$, obtained from the histogram shown in Figure \ref{histsky}.

\clearpage
\section*{Figure Captions}
\figcaption[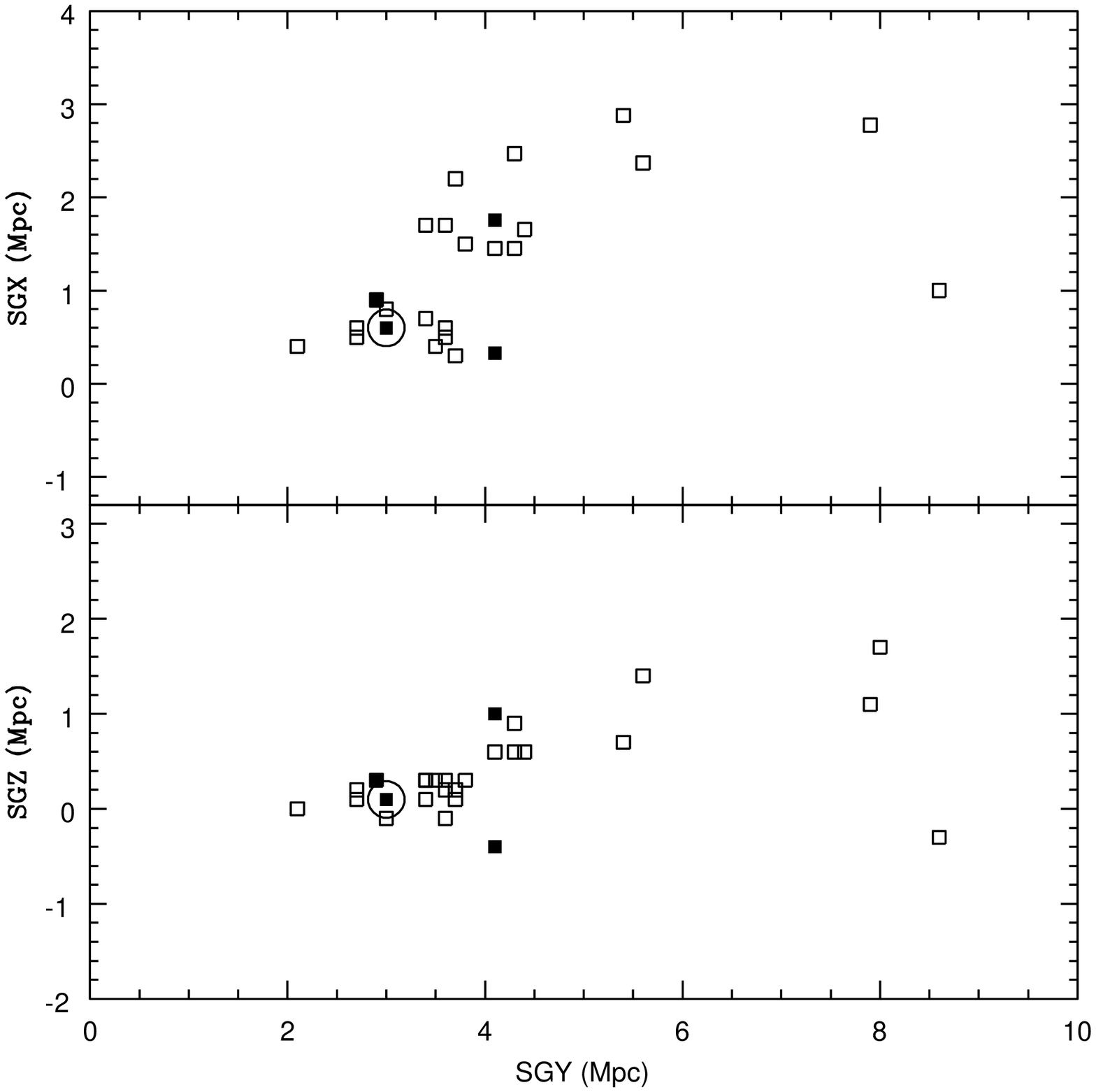]{The environment of NGC 4244: galaxies in the B~4 and B~5 galaxy groups,
plotted in supergalactic coordinates, from Tully 1988. Galaxies brighter than $M_{B, tot}~=~-17.5$
(galaxies which are at least as bright as NGC~4244) are shown by solid symbols, less luminous 
galaxies by open symbols. Total range in luminosity is from $M_{B, tot}$~=~$-$13.0 to $-$19.4.
We have circled NGC~4244.\label{enviro}}
\figcaption[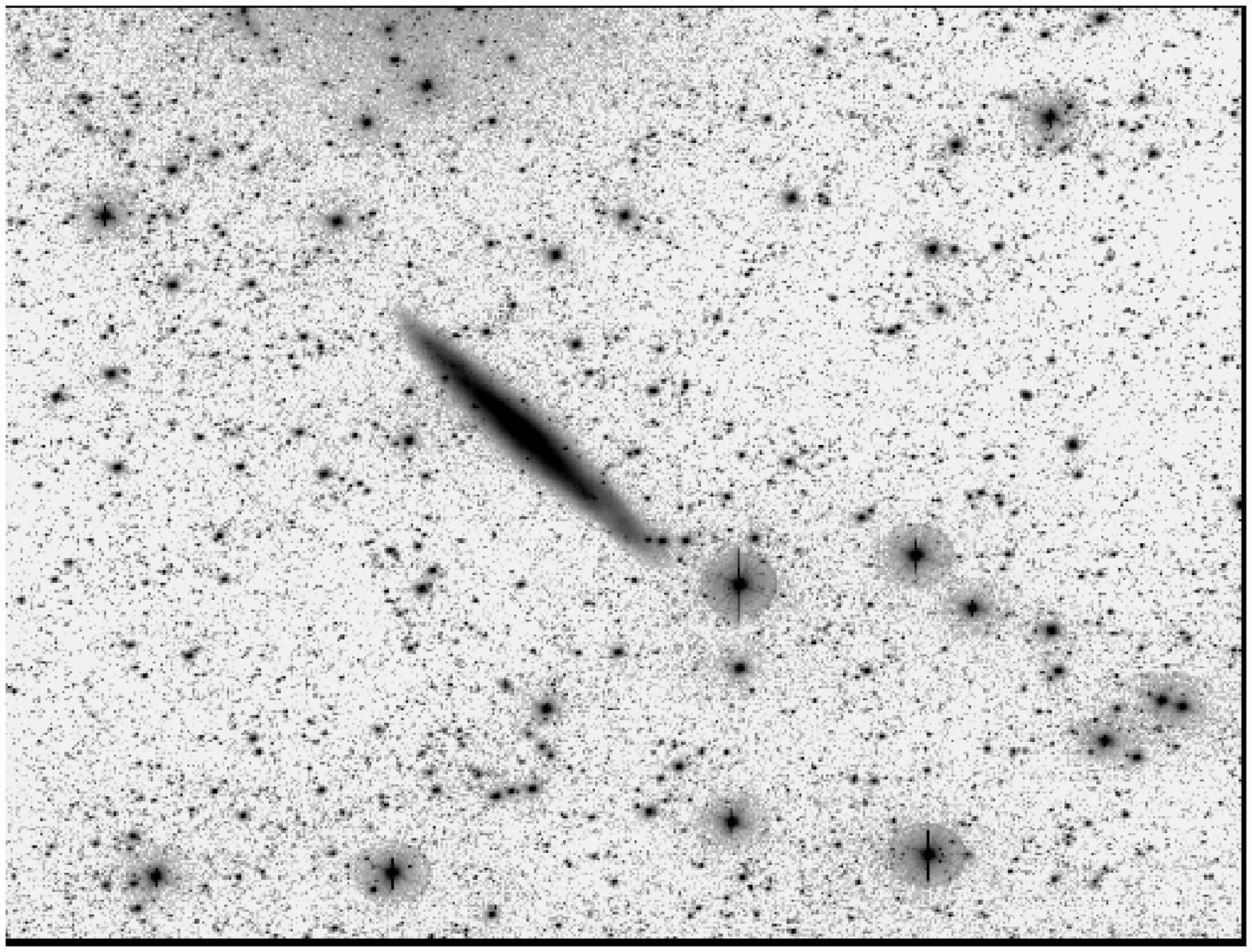]{Our final galaxy image, formed from 70 10-minute 
exposures. The orientation is N is up, E is to the left. The field shown here is 35\arcmin by 30\arcmin.
 \label{final}}
\figcaption[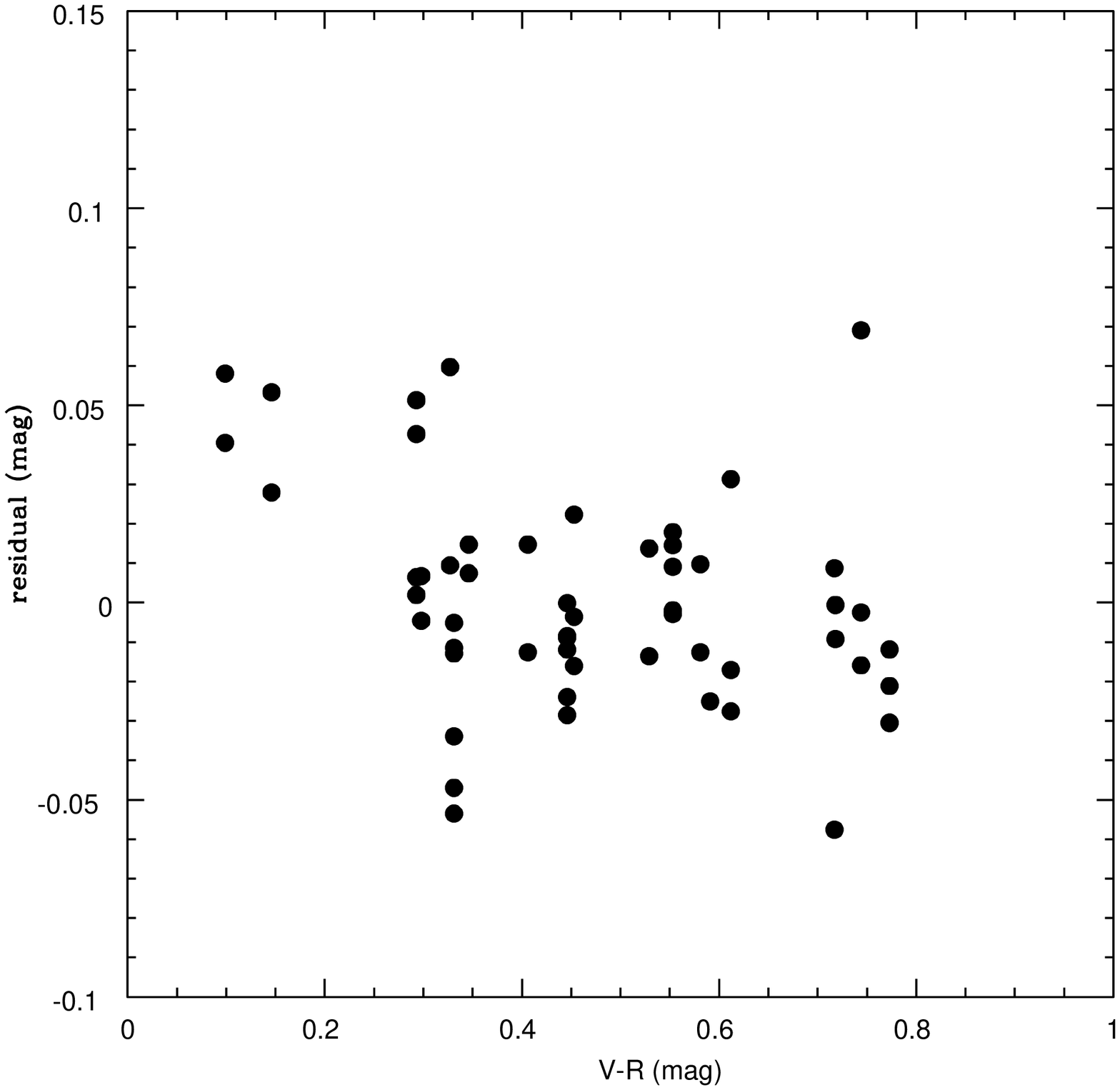]{Residuals from our photometric calibration are plotted as a function
of $V-R$ color of the standard star. Since we observed in only one filter ($R$), we are unable
to correct for a color term in our calibration, but the plot shows that the effect of neglecting
the color term is less than 0.10 mag over the whole range of stellar colors.\label{stds}}
\figcaption[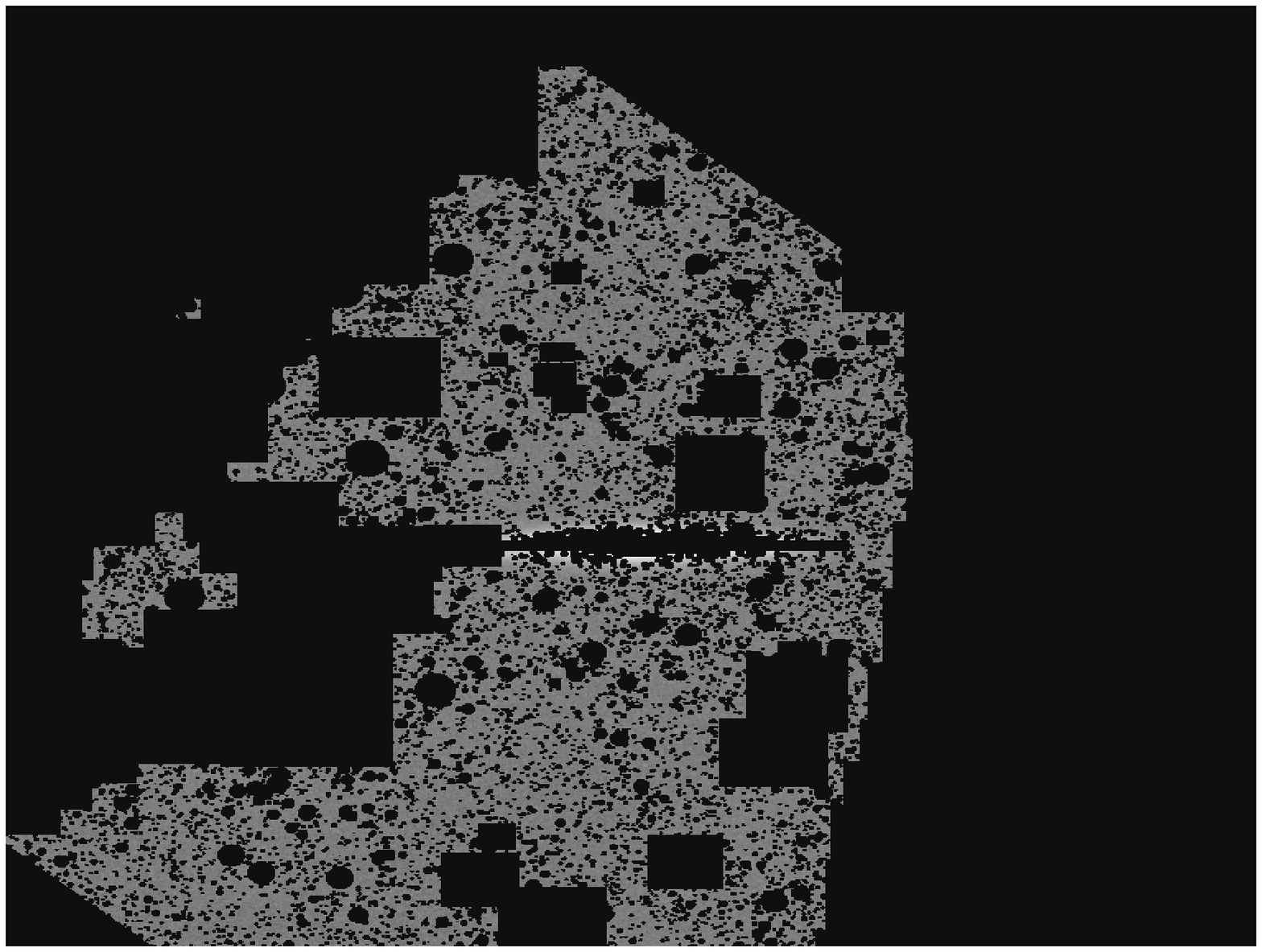]{The galaxy image, rotated counter-clockwise by 42$^{\circ}$ so that the major
axis is parallel to the x axis, with our mask superimposed. The left side of the figure is the 
southwestern side and the right is the northeastern side.\label{maskd}}
\figcaption[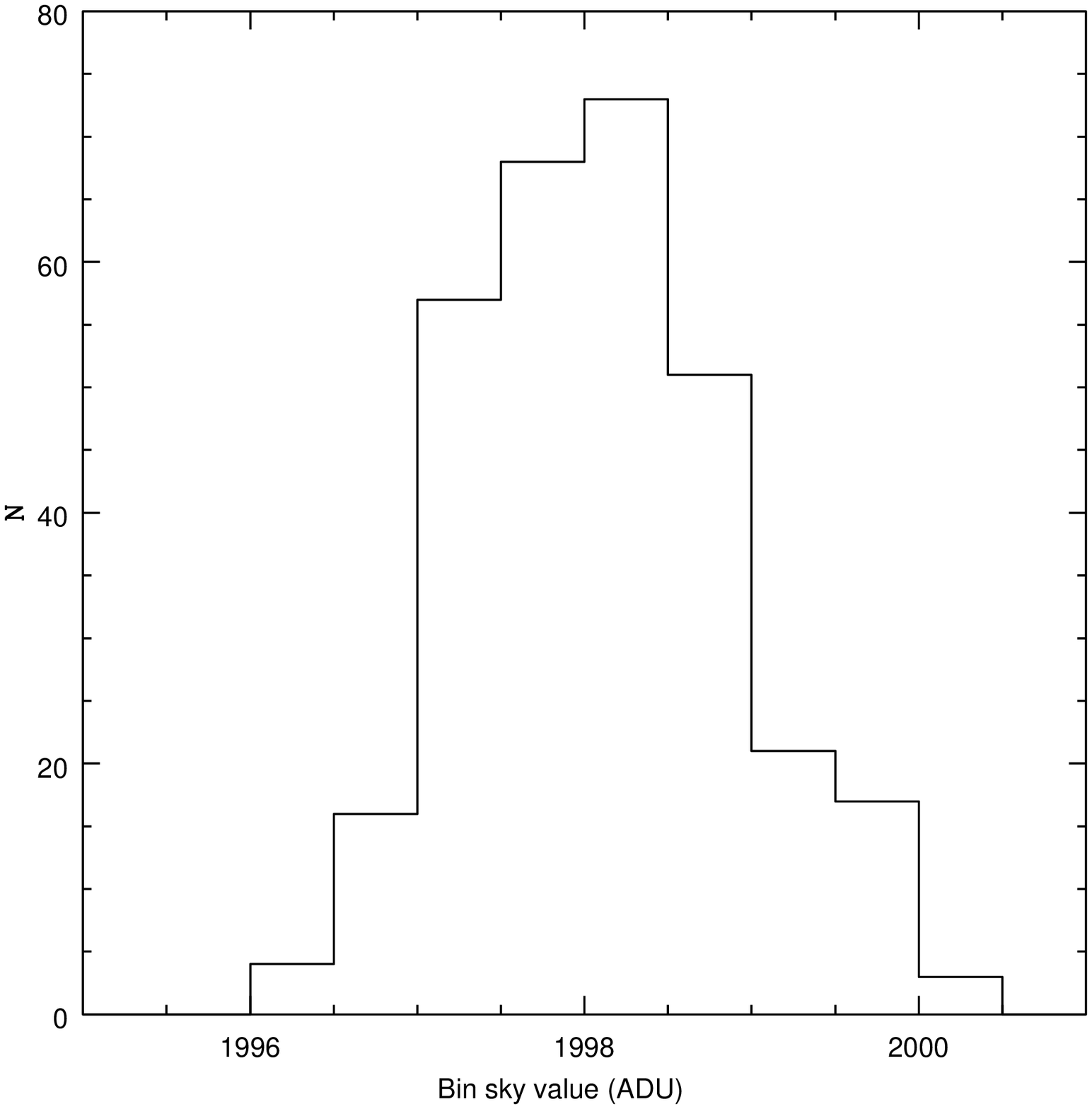]{Histogram of sky values from 310 large bins, located away 
from the galaxy. \label{histsky}}
\figcaption[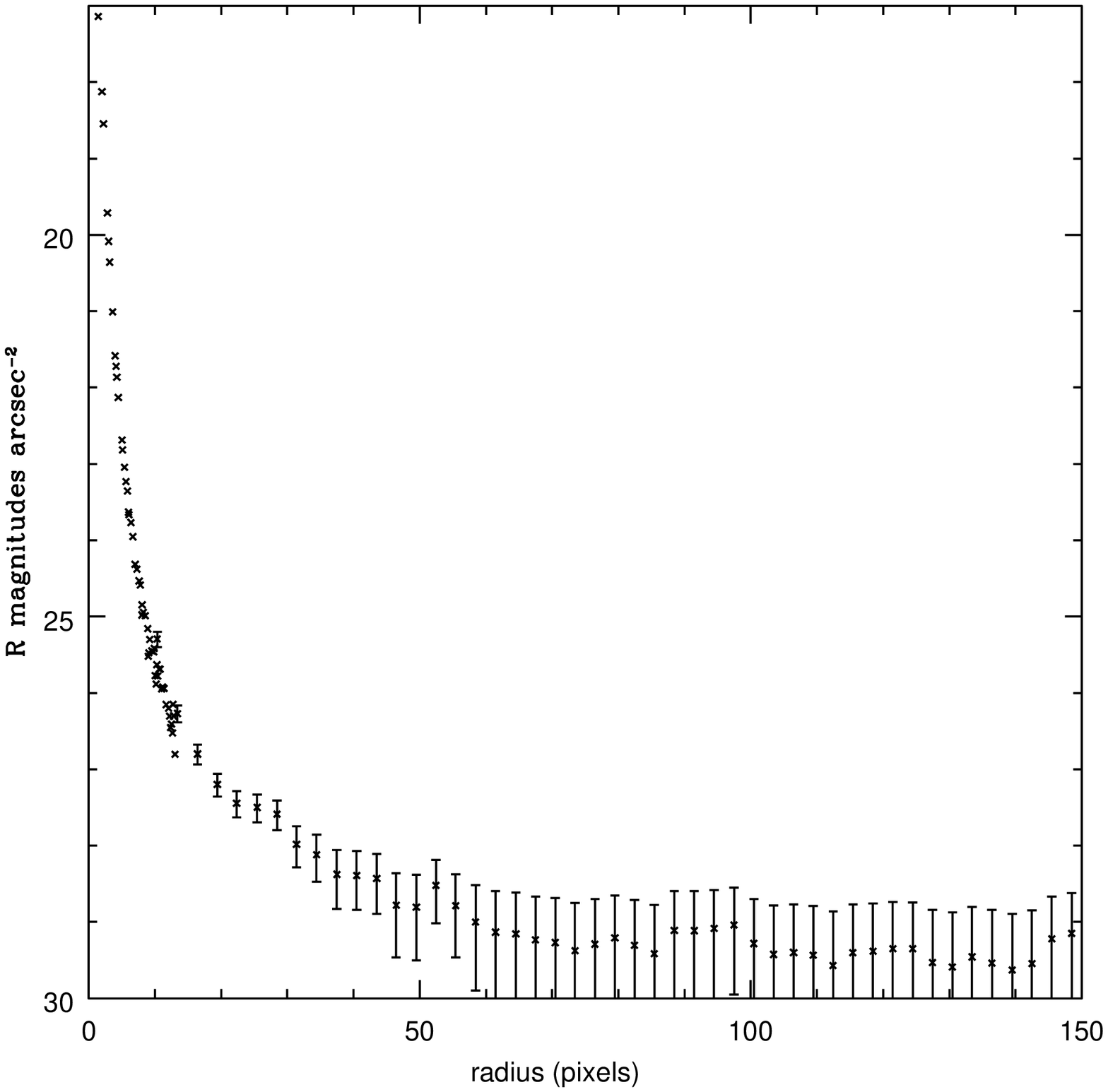]{The stellar PSF, normalized to have the same flux as the central region
of NGC~4244. At a radius of 2~kpc (56 pixels) the scattered light from the center of the galaxy
contributes 28.5 magnitudes arcsec$^{-2}$. \label{psf}} 
\figcaption[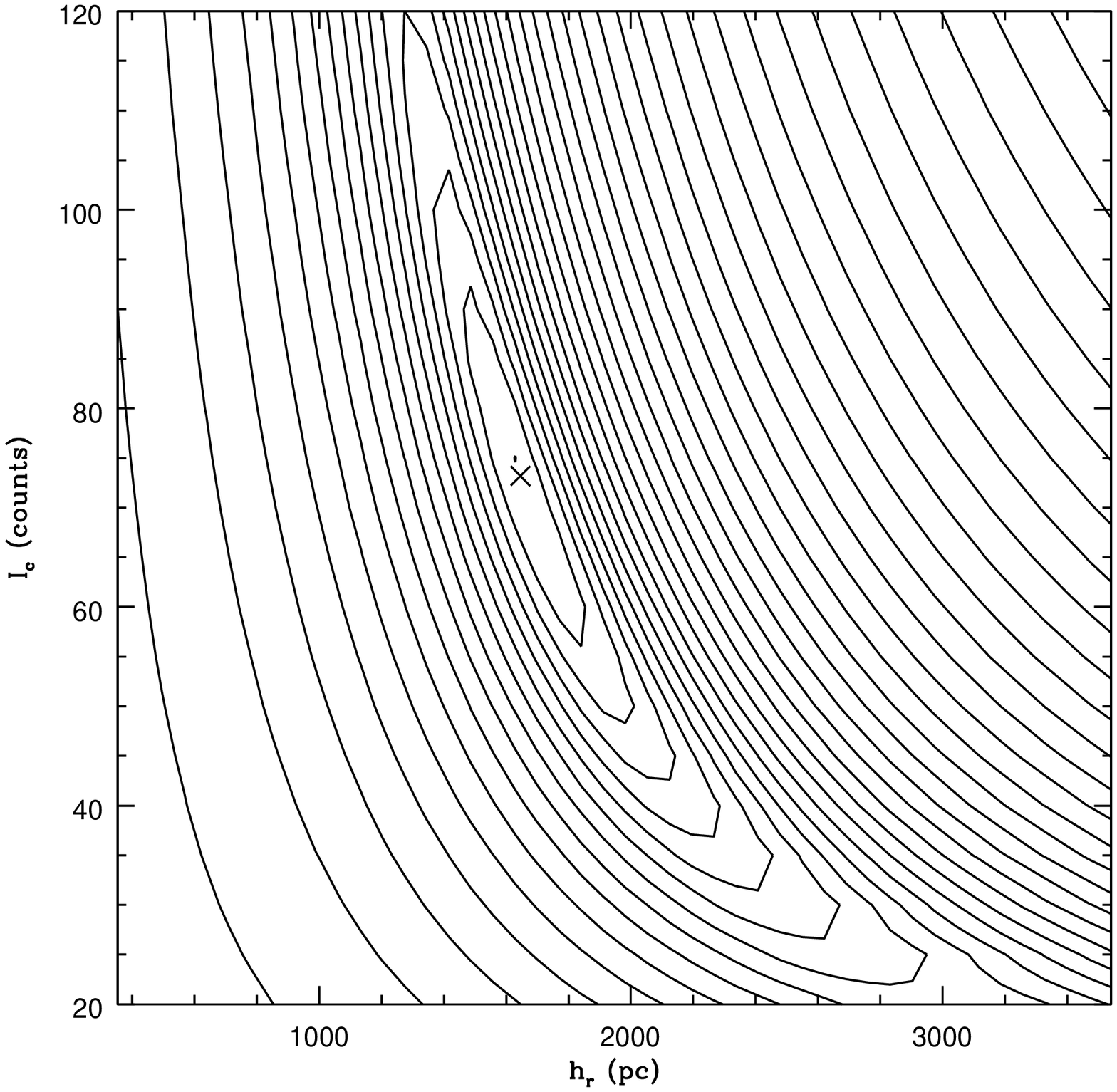]{An example of a contour plot of starting $\chi^2$ values for fixed 
$h_Z$ = 6.6 pixels (233 pc). Contours are logarithmically spaced, with an interval
of 0.1. The final estimate for $h_R$ and $I_C$ is shown with a cross. \label{trench}}
\figcaption[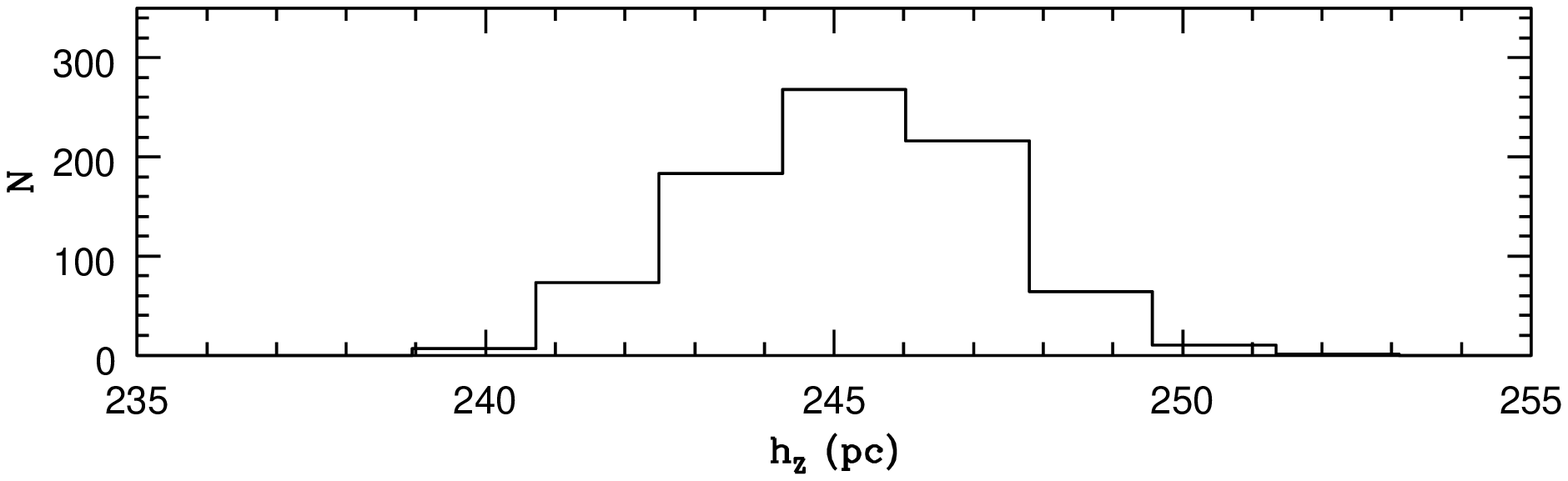]{Histogram of $h_Z$ values estimated by our bootstrap
code. The $\sigma$ measured from these data gives our error estimate for $h_Z$.\label{hist}}
\figcaption[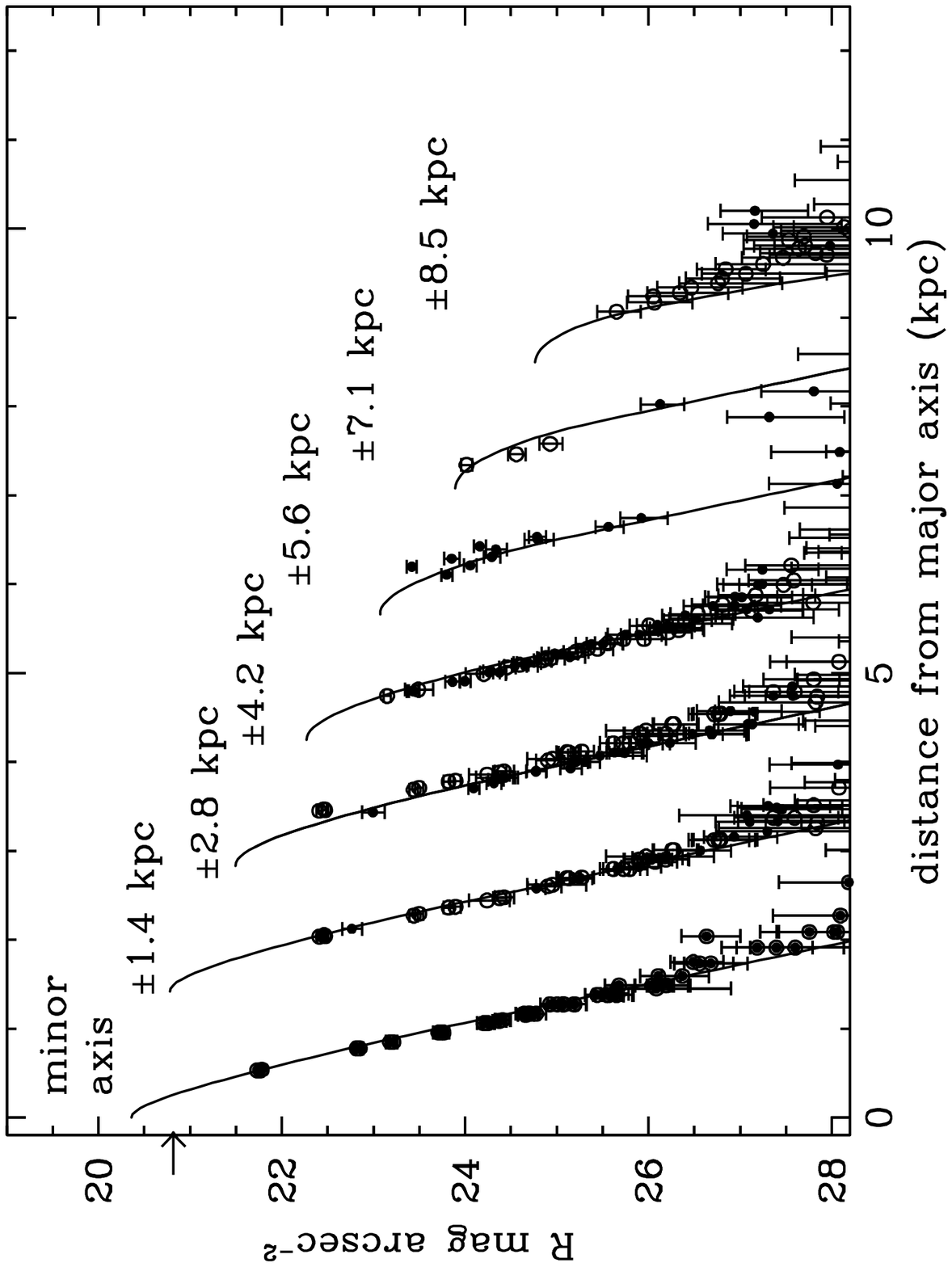]{1-D vertical profiles of NGC~4244, with our best fit exponential
disk model shown as a solid line. The mean sky brightness of $R$ = 20.9 is marked with
an arrow. Adjacent profiles have been shifted by 1.4 kpc for clarity. Open circles are used for
data to the right of the minor axis, filled for data to the left.\label{profiles}}
\figcaption[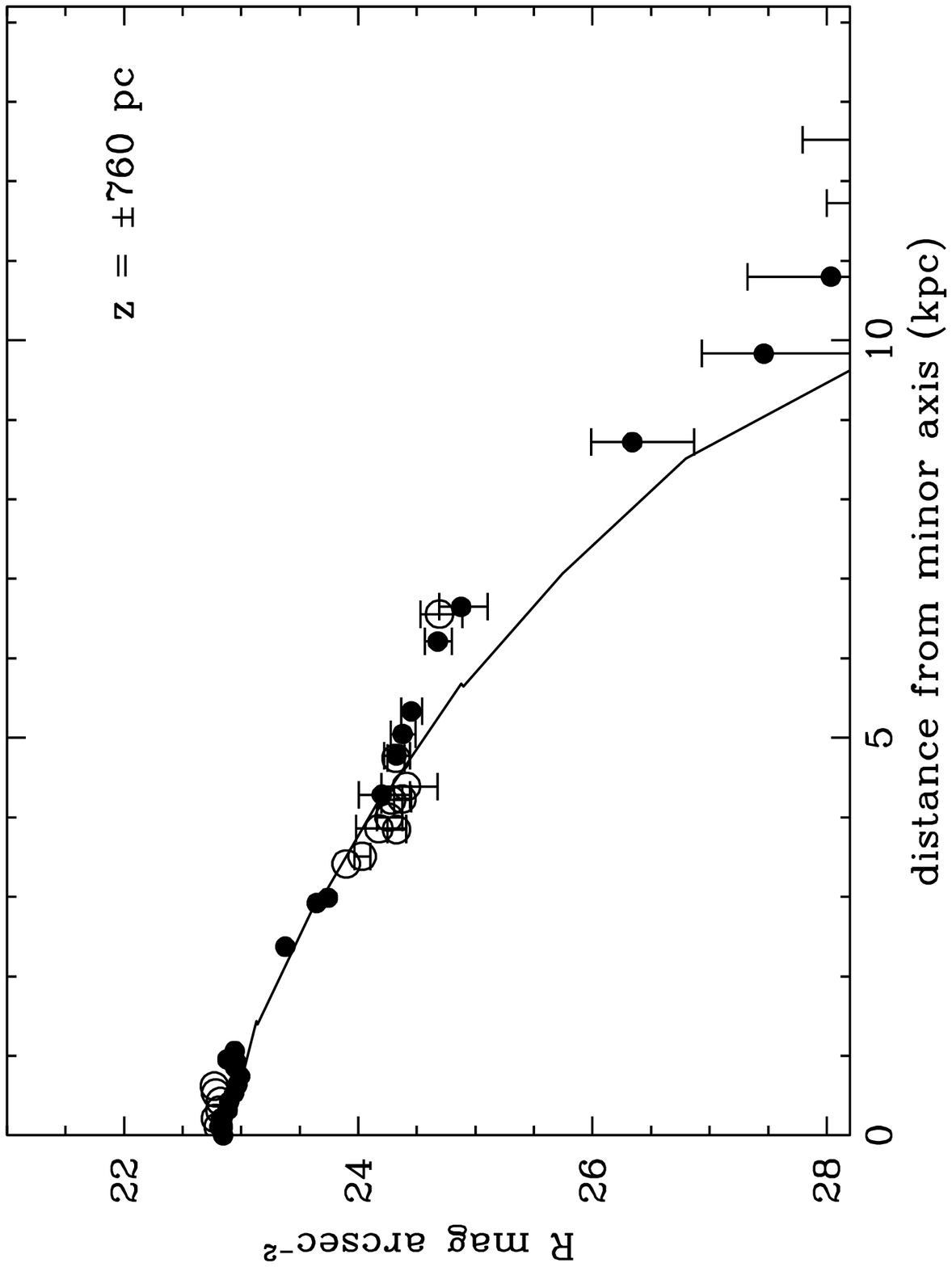]{1-D horizontal profile of NGC~4244, at $\pm$ 760 pc from the major axis. 
Points to the left of the minor axis on Figure 4 are shown by open circles and points to the right
of the minor axis are shown by closed circles. The point at 9.5~kpc appears lower than it 
should, because it straddles the edge of the galaxy and so includes a section outside the
disk cutoff. \label{xp}}
\figcaption[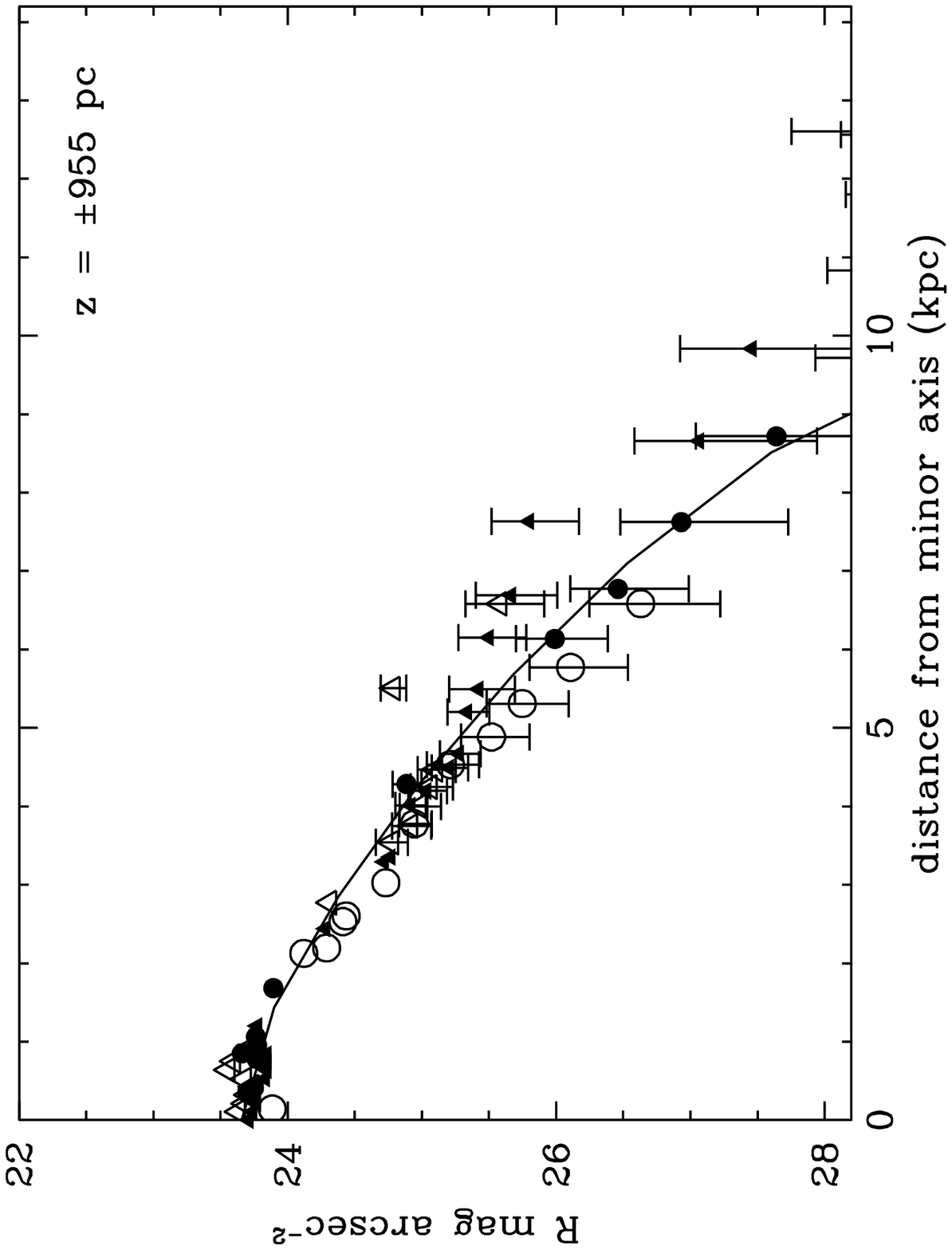]{1-D horizontal profile of NGC~4244, at $\pm$ 955~pc from the major axis. 
All four quadrants of Figure 4 are plotted using different symbols: closed circles and triangles
are left of the major axis, open to the right, triangles below the major axis and circles
above. The warp is clearly visible by $\sim$5~kpc from the minor axis. The full line is the 
model. \label{xp2}}
\figcaption[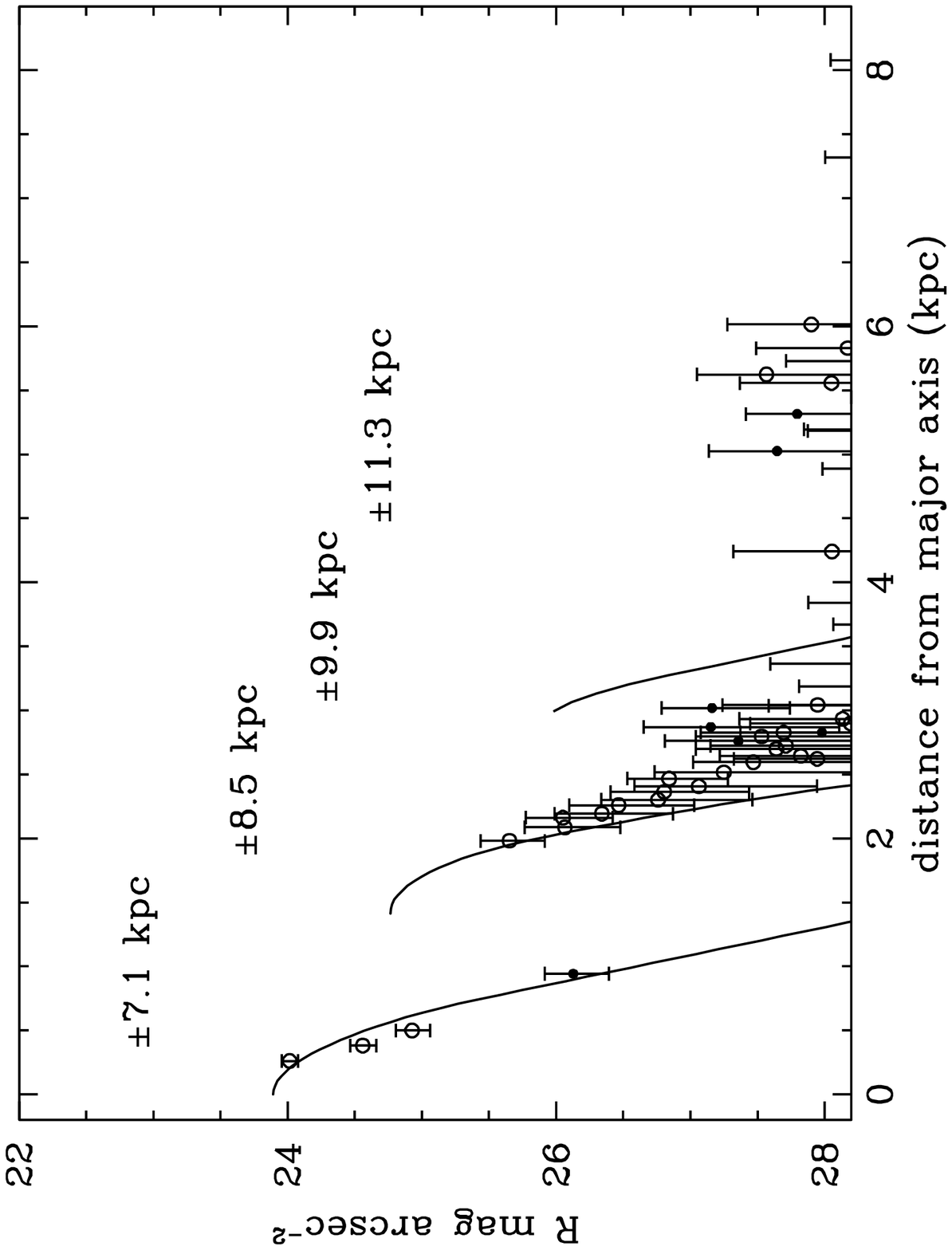]{1-D vertical profiles of the outer disk, showing the disk 
cutoff at $R_{max}$ = 10 kpc. The filled points lie within $R_{max}$ and the open points 
outside of $R_{max}$. By 9.9~kpc, the warp has a large enough effect that points at 
9.9~kpc do not lie on the model line. Open circles are used for
data to the right of the minor axis, filled for data to the left.\label{cutoff}}

\clearpage
\singlespace

\begin{deluxetable}{l c c}
\tablewidth{3.0in}
\singlespace
\tablecaption{Errors in a bin with 500 ADU from galaxy \& 1998 ADU from sky}
\tablenum{1}
\tablehead{\colhead{Source} & \colhead{ADU} & \colhead{\%}}
\startdata
readout noise & 0.03 & 0.00 \nl
Poisson stats & 0.75 & 0.03 \nl
small-scale flat fielding & 0.10 & 0.00 \nl
large-scale flat fielding & 0.60 & 0.02 \nl
surface brightness & 4.60 & 0.19 \nl
\hspace{0.2in} fluctuations \nl
large scale sky variations & 0.80 & 0.03 \nl
\hspace{0.2in} + sky determination \nl
\enddata
\end{deluxetable}

\begin{deluxetable}{l l c c c c c c c c}
\tablecaption{Results of the 2-D Fitting. Parameters with errors quoted were 
estimated, the rest fixed.}
\tablenum{2}
\singlespace
\tablewidth{6.5in}
\tablehead{&&\multicolumn{5}{c}{Correlation coefficients} \\ 
    \colhead{Parameter} & \colhead{Value} & \colhead{$h_R$}&
    \colhead{$h_Z$}& \colhead{$\mu_0$}&\colhead{inc.}&\colhead{$R_{max}$}
    }
\startdata
$h_R$ & 1.84 $\pm$ 0.02 kpc&1.000& 0.442& -0.678& -0.338& -0.465\nl
$h_Z$ & 246 $\pm$ 2 pc&&1.000& -0.512& -0.520& -0.110\nl
$\mu_0$ (observed)& 20.47 $\pm$ 0.05 $R$ magnitudes arcsec$^{-2}$&&&1.000& -0.228& 0.306\nl
inc.& 84.$^{\circ}$5&&&& 1.000& -0.039\nl
$R_{max}$ & 10.0 kpc&&&&&1.000\nl
\hline\nl
$\mu_0$ (face-on)& 22.07 $R$ magnitudes arcsec$^{-2}$\nl
$\mu_0$ (edge-on)& 19.90 $R$ magnitudes arcsec$^{-2}$\nl
\enddata
\end{deluxetable}

\begin{deluxetable}{l c c c l}
\tablecaption{Comparison to Previous Studies}
\tablenum{3}
\tablewidth{6.5in}
\tablehead{\colhead{Author}&\colhead{$h_Z$ (pc)}
           & \colhead{$h_R$ (kpc)}&\colhead{$R_{max}$ (kpc)}&\colhead{$\mu_0$ (magnitude arcsec$^{-2}$)}}
\startdata
van der Kruit \& Searle (1981a)\tablenotemark{a}& 209 & 1.87& 9.91 & 20.6 ($B_J$, edge-on, sech$^2$)\nl
&&&& $\approx$ 19.9 ($B_J$, edge-on, exponential)\nl
&&&& $\approx$ 20.1 ($B$, edge-on)\nl
&&&& $\approx$ 18.6 ($R$, edge-on)\nl
Kodaira \& Yamashita (1996) & 158 & 3.15 & -- & --\nl 
Olling's (1996) reanalysis& 210 & 2 & 10 &21.6 $\pm$ 0.07 ($B$, face-on)\nl 
~~of van der Kruit \& Searle (1981)&&&&$\approx$ 17.6 $\pm$ 0.07 ($R$, edge-on)\nl
\nl
This study& 246 $\pm$ 2 & 1.84 $\pm$ 0.02 & 10.0 & 20.47 $\pm$ 0.05 ($R$, observed)\nl
 & & & & 19.90 ($R$, edge-on)\nl
 & & & & 22.07 ($R$, face-on)\nl
\enddata
\tablenotetext{a}{Revised for our adopted distance of 3.6 Mpc}
\end{deluxetable}

\begin{deluxetable}{l c c}
\tablecaption{Comparison of Thin Disk Parameters between NGC~4244, NGC~5907, \& the Milky Way}
\tablenum{4}
\tablewidth{4.5in}
\tablehead{\colhead{Galaxy}&\colhead{$h_Z$ (pc)} 
           & \colhead{$R_{max}/h_R$}}
\startdata
NGC 4244 & 246 & 5.4 \nl
NGC 5907 & 470 & 3.8 \nl
Milky Way\tablenotemark{a} & 300 & 5.7 \nl 
\enddata
\tablenotetext{a}{References for Milky Way values can be found in Morrison et al. (1997)}
\end{deluxetable}

\begin{figure}
\plotone{Fry.fig1.ps}
\end{figure}
\begin{figure}
\plotone{Fry.fig2.ps}
\end{figure}
\begin{figure}
\plotone{Fry.fig3.ps}
\end{figure}
\begin{figure}
\plotone{Fry.fig4.ps}
\end{figure}
\begin{figure}
\plotone{Fry.fig5.ps}
\end{figure}
\begin{figure}
\plotone{Fry.fig6.ps}
\end{figure}
\begin{figure}
\plotone{Fry.fig7.ps}
\end{figure}
\begin{figure}
\plotone{Fry.fig8.ps}
\end{figure}
\begin{figure}
\plotone{Fry.fig9.ps}
\end{figure}
\begin{figure}
\plotone{Fry.fig10.ps}
\end{figure}
\begin{figure}
\plotone{Fry.fig11.ps}
\end{figure}
\begin{figure}
\plotone{Fry.fig12.ps}
\end{figure}


\begin{thebibliography}{dum}
\bibitem[Aaronson et al. 1986]{a86} Aaronson, M., Bothun, G., \& Mould, 
   J. R. 1986, \apj, 302, 536
\bibitem[Baggett et al. 1998]{bbj98} Baggett, W. E., Baggett, S. M., \& 
   Anderson, K. S. J. 1998, \aj, 116, 1626
\bibitem[Binggeli et al. 1988]{bst88} Binggeli, B., Sandage, A., \& Tammann, G. A.
   1988, \araa, 26, 509
\bibitem[Bothun 1992]{bothun92} Bothun, G. D. 1992, \aj, 103, 104
\bibitem[Bottema 1989]{b89} Bottema, R. 1989, \aap, 221, 236
\bibitem[Briggs 1990]{briggs} Briggs, F. H. 1990, \apj, 352, 15
\bibitem[Bronfman et al 1988]{bronf} Bronfman, L., Cohen, R. S., Alvarez, H., May, J.,
   \& Thaddeus, P. 1988, \apj, 324, 248
\bibitem[Combes \& Becquaert 1997]{combes} Combes, F. \& Becquaert, J.-F. 1997, \aap,
   326, 554
\bibitem[Davis et al. 1985]{d85} Davis, M., Efstathiou, G., Frenk, C. S., 
   White, S. D. M. 1985, \apj, 292, 371
\bibitem[de~Grijs \& Peltier (1997)]{dgp} de~Grijs, R. \& Peltier, R. F. 1997, \aap, 320, L21
\bibitem[de Vaucouleurs 1975]{dv75} de Vaucouleurs, G. 1975, \apj, 202, 610
\bibitem[Dressler 1980]{dress80} Dressler, A. 1980, \apj, 236, 351
\bibitem[Edvardsson et al. 1993]{e93} Edvardsson, B., Andersen, J., Gustafsson, B., 
   Lambert, D.L., Nissen, P.E., Tomkin, J. 1993, \aap, 102, 603
\bibitem[Efron 1982]{e82} Efron, B. 1982, The Jackknife, the Bootstrap, and Other 
   Resampling Plans, CBMS-NSF Regional Conference Series in Applied Mathematics, vol 38 
   (Society for Industrial and Applied Mathematics: Philadelphia, PA) 
\bibitem[Efron \& Tibshirani 1993]{et93} Efron, B. \& Tibshirani, R. J. 1993, An Introduction
to the Bootstrap, Monographs on Statistics and Applied Probability 57 (Chapman \& Hall: New York, NY) 
\bibitem[Eggen et al. 1962]{e62} Eggen, O. J., Lynden-Bell, D., \& Sandage, A. R.
   1962, \apj, 136, 748
\bibitem[Freeman 1970]{freeman70} Freeman, K. C. 1970, \apj, 160, 811
\bibitem[Ferguson et al. (1998)]{ferg98} Ferguson, A. M. N., Wyse, R. F. G., Gallagher, 
   J. S.,\& Hunter, D. A. 1998, \apjl, 506, 19
\bibitem[Jahreiss \& Wielen 1983]{jw83}  Jahreiss, H. \& Wielen, R. 1983 in ``Nearby 
   Stars \& Stellar Luminosity Function'', IAU Colloquium 76, ed. A. G. D. Philip, A. G. 
   Upgren (L. Davis Press: Schenectady, NY), 277
\bibitem[Jenkins \& Binney 1990]{jb90} Jenkins, A. \& Binney, J. 1990, 
   \mnras, 245, 305
\bibitem[Kodaira \& Yamashita (1996)]{ky96} Kodaira, K., \& Yamashita, T. 
   1996, \pasj, 48, 581
\bibitem[Kraan-Kortweg \& Tammann 1979]{kkt79} Kraan-Kortweg, R. C. \&
   Tammann, G. A. 1979, AN, 300, 181
\bibitem[Landolt 1992]{l92} Landolt, A. U. 1992, \aj, 104, 372                
\bibitem[Monkiewicz et al. 1999]{jm98} Monkiewicz, J. A., Morrison, H. L., Harding, P., 
   Boroson, T. A. 1999, in preparation
\bibitem[Morrison et al. (1994)]{mbh} Morrison, H. L., Boroson, T. A., 
   \& Harding, P. 1994, \aj, 108, 1191
\bibitem[Morrison et al. (1997)]{m97} Morrison, H. L., Miller, E. D., 
   Harding, P., Stinebring, D. R., \& Boroson, T. A. 1997, \aj, 113, 2061
\bibitem[Morrison 1999]{m99} Morrison, H. L. 1999, to appear in ``The Third Stromlo
   Symposium: The Galactic Halo,'' ed. B. K. Gibson, T. S. Axelrod, \& M. E. Putnam,
   ASP Conference Series
\bibitem[Olling 1996a]{o96a} Olling, R. 1996, \aj, 112, 457
\bibitem[Olling 1996b]{o96b} Olling, R. 1996, \aj, 112, 481
\bibitem[Olling (1998)]{o98} Olling, R. 1998, private communication
\bibitem[Press et al. 1992]{nr} Press, W. H., Teukolsky, S. A., Vetterling, 
   W. T., \& Flannery, B. P. 1992, Numerical Recipes in Fortran, Second 
   Edition (Cambridge University Press, Cambridge) 
\bibitem[Roberts \& Haynes 1994]{rharaa} Roberts, M. S. \& Haynes, M. P. 1994, \araa, 32, 115
\bibitem[Quinn \& Goodman 1986]{qg86} Quinn, P. J.  \& Goodman, J. 1986, 
   \apj, 309, 472
\bibitem[Sackett et al. 1994]{sack94} Sackett, P. D., Morrison, H. L., Harding, P., 
   Boroson, T. A. 1994, \nat, 370, 441
\bibitem[Sage (1993)]{sage93} Sage, L. J. 1993, \aap, 272, 123
\bibitem[Sarajedini et al. 1998]{ata} Sarajedini, A., Geisler, D., Harding, P., \& Schommer,
R. 1998, \apjl, 508, L37
\bibitem[Schommer et al. (1991)]{schommer} Schommer, R. A., Christian, C. A., Caldwell, N., 
Bothun, G. D., \& Huchra, J. 1991, \aj, 101, 873
\bibitem[Scoville et al. 1993]{sco} Scoville, N. Z., Thakkar, D., Carlstrom, J. E., \&
   Sargent, A. I. 1993, \apjl, 404, 59
\bibitem[Searle \& Zinn 1978]{sz78} Searle, L. \& Zinn, R. 1978, \apj, 225, 357
\bibitem[Silva et al. 1989]{silva89} Silva, D. R., Boroson, T. A., Thompson, I. B.,
   \& Jedrzejewski, R. I. 1989, \aj, 98, 131
\bibitem[Steidel et al. 1996]{stei} Steidel, C. C., Giavalisco, M., Pettini, M., Dickinson, 
   M., \& Adelberger, K. L. 1996, \apj, 462, L17
\bibitem[Stetson 1987]{s87} Stetson, P. B. 1987, \pasp, 99, 191
\bibitem[Tonry \& Schneider (1988)]{ts88} Tonry, J., \& Schneider, D. P. 
   1988, \aj, 96, 807
\bibitem[Tsikoudi 1980]{vt80} Tsikoudi, V. 1980, \apjs, 43, 365
\bibitem[Tully 1988]{tully} Tully, R. B. 1988, Nearby Galaxies Catalog (Cambridge 
   University Press: Cambridge, England) 
\bibitem[van der Kruit \& Searle (1981a)]{vdks} van der Kruit, P. C., \& 
   Searle, L. 1981a, \aap, 95, 105
\bibitem[van der Kruit \& Searle 1981b]{vdks81b} van der Kruit, P. C., \& 
   Searle, L. 1981b, \aap, 95, 116
\bibitem[van der Kruit \& Searle 1982a]{vdks82a} van der Kruit, P. C., \& 
   Searle, L. 1982a, \aap, 110, 61
\bibitem[van der Kruit \& Searle 1982b]{vdks82b} van der Kruit, P. C., \& 
   Searle, L. 1982b, \aap, 110, 79
\bibitem[Wolfe 1990]{wolfe90} Wolfe, A. M. 1990, in ``The Interstellar Medium 
   in Galaxies; Proceedings of the 2nd Teton Conference,'' ed. H. A. Thronson, Jr. \& 
   J. M. Shull (Kluwer Academic Publishers: Dordrecht, Netherlands), 387
\bibitem[Wouterloot et al. 1990]{wouter} Wouterloot, J. G. A., Brand, J., Burton, W. B.,
   \& Kwee, K. K. 1990, \aap, 230, 21
\end{thebibliography}
\end{document}